\documentclass{acmtrans2e}

\usepackage{psfig}

\newcommand{\asfsdf}{ASF+SDF}
\newcommand{\sdf}{SDF}
\newcommand{\asf}{ASF} 
\newcommand{\muasf}{$\mu$ASF}
\newcommand{\muasfplus}{$\mu$ASF$^+$}

\firstfoot{}

\runningfoot{}

\title{Compiling Language Definitions: The \asfsdf\ Compiler}

\author{
M. G. J. VAN DEN BRAND\\ CWI\\
J. HEERING\\ CWI\\
P. KLINT\\ CWI and University of Amsterdam \and
P. A. OLIVIER\\ CWI}

\begin{abstract}
The \asfsdf\ Meta-Environment is an interactive language development
environment whose main application areas are definition of
domain-specific languages, generation of program analysis and
transformation tools, production of software renovation tools, and
general specification and prototyping.  It uses conditional rewrite
rules to define the dynamic semantics and other tool-oriented aspects
of languages, so the effectiveness of the generated tools is
critically dependent on the quality of the rewrite rule
implementation.

The \asfsdf\ rewrite rule compiler generates C code, thus taking
advantage of C's portability and the sophisticated optimization
capabilities of current C compilers as well as avoiding potential
abstract machine interface bottlenecks.  It can handle large
(10\,000$+$ rule) language definitions and uses an efficient run-time
storage scheme capable of handling large (1\,000\,000$+$ node) terms.
Term storage uses maximal subterm sharing (hash-consing), which turns
out to be more effective in the case of \asfsdf\ than in Lisp or SML.
Extensive benchmarking has shown the time and space performance of the
generated code to be as good as or better than that of the best
current rewrite rule and functional language compilers.
\end{abstract}

\category{D.3.1}{Programming Languages}{Formal Definitions and Theory}[Semantics]
\category{D.3.2}{Programming Languages}{Language Classifications}[Specialized application languages]
\category{D.3.4}{Programming Languages}{Processors}[code generation; compilers; optimization]
\category{F.4.2}{Mathematical Logic and Formal Languages}{Rewriting Systems}

\begin{document}

\begin{bottomstuff}
This research was supported in part by the Telematica Instituut under the
{\it Domain-Specific Languages} project.  Parts of this article
emphasizing memory management issues have appeared in preliminary form
in {S. J\"{a}hnichen} (ed.), Compiler Construction (CC '99), vol. 1575
of {\it Lecture Notes in Computer Science}, Springer-Verlag, 1999,
pp. 198--213. \\
Authors' addresses: 
M. G. J. van den Brand, Department of Software Engineering, CWI, 
Kruislaan 413, 1098 SJ Amsterdam, The Netherlands; email: Mark.van.den.Brand@cwi.nl;
J. Heering, Department of Software Engineering, CWI,
Kruislaan 413, 1098 SJ Amsterdam, The Netherlands; email: Jan.Heering@cwi.nl;
P. Klint, Department of Software Engineering, CWI, 
Kruislaan 413, 1098 SJ Amsterdam, The Netherlands; email: Paul.Klint@cwi.nl;
P. A. Olivier, Department of Software Engineering, CWI, 
Kruislaan 413, 1098 SJ Amsterdam, The Netherlands; email: Pieter.Olivier@cwi.nl.
% \permission
% \copyright
\end{bottomstuff}

\maketitle

\section{Introduction} \label{sec:INTRO}

\asfsdf\ \cite{BHK89,DHK96} is the metalanguage
of the \asfsdf\ Meta-Environment \cite{Kli93.meta}, an
interactive environment for the development of domain-specific and
general purpose programming languages, covering parsing, typechecking,
translation, transformation, and execution of programs.

\sdf\ \cite{HHKR89.update}, the syntax definition component of \asfsdf, 
is a BNF-like formalism for defining the lexical, context-free and
abstract syntax of languages.  The implementation of \sdf\ is beyond
the scope of this article.  Suffice it to say, its implementation
supports interactive syntax development and fully general context-free
parsing by means of scanner and parser generators that are both lazy
(just-in-time) and incremental
\cite{HKR90,HKR92.igls,HKR94}. \sdf\ is currently being superseded by
SDF2 \cite{Visser:97}, whose main feature is a very close integration
of lexical and context-free syntax. This is reflected in its
implementation by the use of scannerless parsing.

The semantics definition component of \asfsdf, which is an outgrowth
of the algebraic specification formalism \asf\ \cite{BHK89}, uses
rewrite rules to describe the semantics of languages.  Such semantics
may be static (typechecking) or dynamic.  The latter may have an
interpretive or translational character, it may include program
transformations, and so on. These are all described in terms of
rewrite rules whose left- and right-hand sides are sentences in the
language defined by the \sdf-part of the language definition.

Rewriting is the simplification of algebraic expressions or terms
everybody is familiar with. It is ubiquitous in (computer) algebra as
well as in algebraic semantics and algebraic specification.  It is
also important in functional programming, program transformation and
optimization, and equational theorem proving.  Useful theoretical
surveys of rewriting are \cite{Klo92,DJ90}, but we assume only a basic
understanding of rewrite systems on the part of the reader. In
addition to regular rewrite rules, \asfsdf\ features conditional
rewrite rules, associative (flat) lists, and default rules. These will
be explained below.

\asfsdf\ is more expressive than attribute grammars, which it
includes as the subclass of definitions that are non-circular
primitive recursive schemes (NPRSs) \cite{CF82}. This is the natural
style for most typecheckers and translators.  Using this
correspondence, \citeN{Meu96} has transferred incremental evaluation
methods originally developed for attribute grammars to NPRS-style
\asfsdf\ definitions.
  
\asfsdf's main application areas are 
\begin{itemize}
\item Definition of domain-specific languages 

\item Generation of program analysis and
      transformation tools

\item Production of software renovation tools

\item General specification and prototyping.  
\end{itemize}
Table~\ref{table:APPS} gives details and further references.

\begin{table}
\begin{center}
\begin{tabular}{|p{12cm}|} \hline

\textbf{Domain-Specific Languages}\\
\hline
\begin{itemize}
\item Risla \cite{BDKKM96,DK98} (financial product specification)

\item Box \cite{BV96} (prettyprinting)

\item EURIS \cite{GKV95} (railroad safety)

\item Action Semantics \cite{Deu94.thesis} (programming language semantics)

\item Dahl \cite{Moo97} (dataflow analysis)

\item Manifold \cite{RT92}, ToolBus \cite{TB-SCP98} (coordination languages)

\item ALMA-0 \cite{AptEtAl:98} (backtracking and search)
\end{itemize}
\\ \hline\hline

\textbf{Program Analysis}\\
\hline
\begin{itemize}
\item Typechecking of Pascal \cite[Chapter 2]{DHK96}

\item Typechecking and execution of CLaX \cite{DT92.anim,DT97.slice}

\item Type inference, object identification, and
      documentation for Cobol \cite{BSV00,DM98.types,DK98.iwpc,DK99}
\end{itemize}
\\ \hline\hline

\textbf{Program Transformation}\\
\hline
\begin{itemize}
\item Interactive program transformation for 
        Clean \cite{BEGMOP95} and Prolog \cite{Bru96}

\item Automatic program transformation for
        C++ \cite{DineshEtAl:98} 
\end{itemize}
\\ \hline\hline

\textbf{Software Renovation}\\
\hline
\begin{itemize}
\item Description of the multiplicity of languages and dialects
      encountered in software renovation applications
      such as Cobol (including embedded languages like SQL and CICS)
      \cite{BKV96b,BKV97b,DKV99}

\item Automatic program transformation for restructuring of 
        Cobol programs (including embedded languages like SQL and CICS) 
        \cite{BSV97a,BSV98.goto,SSV99}

\item Derivation of language descriptions from compilers and on-line manuals
      \cite{SV99,SellinkVerhoef:00}
\end{itemize}
\\ \hline\hline

\textbf{Specification and Prototyping of New Applications and Tools}\\
\hline
\begin{itemize} 
\item PIM \cite{Fie92,BDFH97} (compiler toolkit)

\item $\mu$CRL \cite{Hil96} (proof checking and simulation toolkit)
 
\item Components of the \asfsdf\ Meta-Environment itself \cite{BKMO97}
      (including a parser generator, a prettyprinter generator, and
      the \asfsdf\ compiler described in this article)
\end{itemize}
\\ \hline

\end{tabular}
\end{center}
\caption{\label{table:APPS} Main application areas of the \asfsdf\ Meta-Environment.}
\end{table}

The effectiveness of the tools generated by the \asfsdf\ Meta-Environment is
critically dependent on the quality of the rewriting implementation.
The original interpretive implementation left room for improvement.
Its author, inspired by earlier rewrite compilation work of
\citeN{Kaplan:87}, sketched a more efficient compilational scheme
\cite{Dik:89} that ultimately served as a basis for the compiler described here.

We describe the current \asfsdf\ compiler and compare its performance
with that of other rewrite system and functional language
compilers we were able to run, namely, 
Clean \cite{PlasmeijerVanEekelen:94,SmetsersEtAL:91}, 
Elan \cite{MoreauKirchner:98}, 
Haskell \cite{PeytonJoneEtAl:93,PeytonJones:96}, 
Opal \cite{DidrichEtAl:94},
and SML \cite{Appel:92}.

The real-world character of \asfsdf\ applications has important
consequences for the compiler:
\begin{itemize}

\item It must be able to handle \asfsdf\ definitions of up to
      50\,000 lines. Disregarding layout and syntax
      declarations (\sdf-parts), this corresponds to 10\,000
      (conditional) rewrite rules.

\item It must include optimizations for the major sources of
      inefficiency encountered in practice. 

\item It has to support separate compilation of \asfsdf\ modules.
      For large language definitions, modularization and separate
      compilation are as important as for conventional programs.

\end{itemize}

This article is organized as follows: 
general compilation scheme (Sec.~\ref{sec:GENSCHEME});
major design considerations (Sec.~\ref{sec:DESIGN});
the \asfsdf\ language (Sec.~\ref{sec:ASFSDF});
preprocessing (Sec.~\ref{sec:PREPROC});
code generation (Sec.~\ref{sec:CODEGEN});
postprocessing (Sec.~\ref{sec:POSTPROC});
benchmarking (Sec.~\ref{sec:BM});
conclusions and further work (Sec.~\ref{sec:CONC}).
Related work is discussed at
appropriate points throughout the text rather than in a separate
section.

\section{General Compilation Scheme} \label{sec:GENSCHEME}

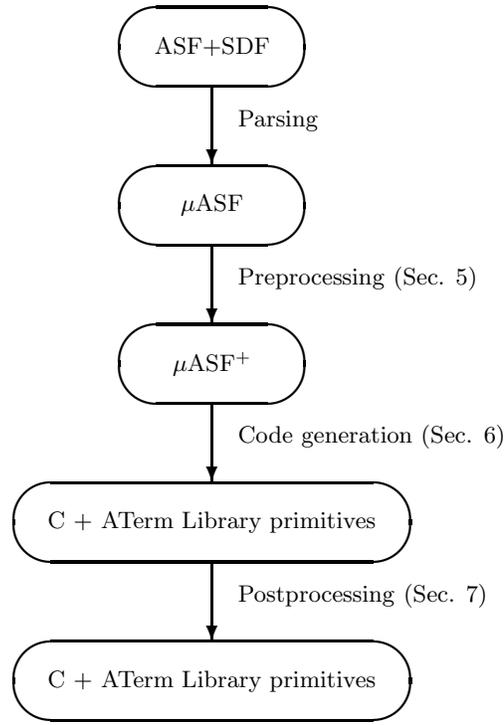
\begin{figure}
\rule{12.5cm}{.5mm}
\small

\begin{picture}(400,280)(-150,-260)
% last tuple = position

\thicklines

\put(0,0)
  {\begin{picture}(200,200)
      \put(0,0){\oval(70,30)}
      \put(0,0){\makebox(0,0){\asfsdf}}
      \put(0,-15){\vector(0,-1){30}}
      \put(10,-30){Parsing}
  \end{picture}
  }

\put(0,-60)
  {\begin{picture}(200,200)
      \put(0,0){\oval(70,30)}
      \put(0,0){\makebox(0,0){\muasf}}
      \put(0,-15){\vector(0,-1){30}}
      \put(10,-30){\shortstack[l]{
                    Preprocessing (Sec. \ref{sec:PREPROC})
                    }
                   }      
  \end{picture}
  }

\put(0,-120)
  {\begin{picture}(200,200)
      \put(0,0){\oval(70,30)}
      \put(0,0){\makebox(0,0){\muasfplus}}
      \put(0,-15){\vector(0,-1){30}}
      \put(10,-30){\shortstack[l]{
                    Code generation (Sec. \ref{sec:CODEGEN})
                    }
                  }         
  \end{picture}
  }

\put(0,-180)
  {\begin{picture}(200,200)
      \put(0,0)  {\oval(150,30)}
      \put(0,0)  {\makebox(0,0){C $+$ ATerm Library primitives}}
      \put(0,-15){\vector(0,-1){30}}
      \put(10,-30){\shortstack[l]{
                    Postprocessing (Sec. \ref{sec:POSTPROC})
                    }
                  }
  \end{picture}
  }

\put(0,-240)
  {\begin{picture}(200,200)
      \put(0,0){\oval(150,30)}
      \put(0,0){\makebox(0,0){C $+$ ATerm Library primitives}}
  \end{picture}
  }
\end{picture}
\normalsize
\rule{12.5cm}{.5mm}
\caption{General layout of the \asfsdf\ compiler.} \label{fig:CompLayoutRough}
\end{figure}

Before we discuss the major design issues, it is useful for the reader
to understand the general layout of the compiler as shown in Figure
\ref{fig:CompLayoutRough}. The following compiler phases can be distinguished:
\begin{itemize}

\item Parsing. Since the syntax of \asfsdf-definitions is largely 
      defined by their \sdf-part, parsing them is a nontrivial two-pass
      process, which is beyond the scope of this article. Suffice it
      to say, this phase yields an abstract syntax representation of
      the input definition as usual.  As indicated in the second
      box from the top, the parser's output formalism is \muasf, an
      abstract syntax version of \asfsdf.
%     We don't mention Asfix to avoid undue complication.

\item Preprocessing. This is performed on the \muasf\
      representation, which is very close to the source level.
      Typical examples are detection of variable bindings
      (``assignments'') in conditions and introduction of
      {\tt else}s for pairs of conditional rewrite rules with
      identical left-hand sides and complementary conditions. The
      output formalism of this phase is \muasfplus, a superset of
      \muasf.

\item Code generation. The compiler generates C extended
      with calls to the \emph{ATerm library}, a run-time library for
      term manipulation and storage. Each \muasf\ function is compiled
      to a separate C function.  The right-hand side of a rewrite rule
      is translated directly to function calls if necessary.  Term
      matching is compiled to a finite automaton. List matching code
      depends on the complexity of the pattern involved. A few special
      list patterns that do not need backtracking are eliminated by
      transforming them to equivalent term patterns in the
      preprocessing phase, but the majority is compiled to special
      code.
 
\item Postprocessing. This is performed on the C code generated in
      the previous phase. A typical example is constant caching.

\end{itemize}

\section{Major Design Considerations} \label{sec:DESIGN}

The design of the compiler was influenced by the experience gained in
previous compiler activities within the \asfsdf\ project itself
\cite{Dik:89,FokkinkEtAl:98,Hen91,Kamperman:96,Walters:97} as well as
in various functional language and Prolog compiler projects
elsewhere. The surveys \cite{HartelEtAl:96} on functional language
compilation and \cite{VanRoy:93} on Prolog compilation were
particularly helpful.

In the following subsections we discuss the arguments in favor of
generating C rather than native code, the choice of \asfsdf\ as an
implementation language for the compiler, some pitfalls in the areas
of high-level transformations and abstract machine interfaces, the
importance of a proper organization of term storage, and some issues
related to separate compilation.

\subsection{Choice of C as Target Language}

  Generating C code is an efficient way to achieve portability.  Folk
  wisdom has it that C code is 2--3 times slower than native code, but
  this is not borne out by the ``Pseudoknot'' benchmark results
  reported in \cite[Table~9]{HartelEtAl:96}, where the best functional
  language and rewrite system compilers generate C code.  The probable
  reason is that many C compilers perform sophisticated optimizations
  \cite{Muchnick:97}, although this raises the issue of tuning the
  generated C code to the optimizations done by different C compilers.
  At least in our case, the fact that C is in some respects less than
  ideal as a compiler target \cite{PeytonJonesEtAl:98} does not
  invalidate these favorable observations.

\subsection{Choice of \asfsdf\ as Implementation Language}

  Not unexpectedly, large parts of the compiler can be expressed very
  naturally in \asfsdf, so it was decided to write the compiler in its
  own source language. Since the compiler is fairly large,
  self-compilation is an interesting benchmark.

\subsection{Pitfalls in High-Level Transformations and Abstract Machine
            Interfaces---The Bottleneck Effect} \label{sec:PITFALLS}

  High-level transformations have to be applied with extreme care,
  especially if their purpose is to simplify the compiler by reducing
  the number of different constructs that have to be handled later on.
  For instance, by first transforming conditional rewrite rules to
  unconditional ones or associative list matching to term matching,
  the compiler can be simplified considerably, but at the expense of a
  serious degradation in the performance of the generated code.
  Similarly, transformation of default rules (which can be applied
  only when all other rules fail) to sets of ordinary rewrite rules
  that catch the same cases would lead to very inefficient code.
  These transformations would perhaps be appropriate in a formal
  semantics of \asfsdf, but in a compiler they cause a bottleneck
  whose effect is hard to undo at a later stage.

  For this reason, our compiler does not generate code for the
  Abstract Rewrite Machine (ARM), which was originally developed for
  \asfsdf\ and then used in the compiler for the equational
  programming language Epic \cite{FokkinkEtAl:98}. ARM is based on the
  notion of \emph{minimal term rewriting system} (MTRS).  An MTRS
  consists of unconditional rewrite rules in so-called \emph{minimal
  form} \cite[Definition~3.1.1]{FokkinkEtAl:98}.  ARM thus requires a
  high-level transformation phase to simplify the rules that are not
  in this form and to eliminate the conditions (if any)
  \cite[p.~681]{FokkinkEtAl:98}.  Furthermore, ARM does not support
  list matching, so rules with lists have to be transformed to minimal
  rewrite rules as well. Although these transformations are possible,
  they have turned out to be counterproductive in the \asfsdf\
  compiler, and with C taking care of portability, ARM's main purpose
  was lost. In fact, rather than breaking rules down into smaller
  ones, the \asfsdf\ compiler tries to combine rules into larger ones
  as much as possible during preprocessing.

  Our experience with ARM is not unique. Any fixed abstract machine
  interface is a potential bottleneck in the compilation
  process. The modularization advantage gained by
  introducing it may be offset by a serious loss in opportunities for
  generating efficient code.  The factors involved in this trade-off
  have a qualitatively different character.  The abstract machine
  interface facilitates \emph{construction} and \emph{verification} of
  the compiler, but possibly at the expense of the \emph{performance
  of the generated code}.  See the instructive discussion in
  \citeN[Sec.~2.4]{VanRoy:93} on the pros and cons of the use of the
  Warren Abstract Machine (WAM) in Prolog compilers. Although the
  bottleneck effect is hard to describe in quantitative terms, it has
  to be taken seriously, the more so since the elegance of the
  abstract machine approach is not conducive to a thorough analysis of
  its performance in terms of overall compiler quality. 

  Of course, C also acts as an abstract machine interface, but,
  compared with ARM or other abstract machines, it is much less
  specialized and more flexible, acting proportionally less as a
  bottleneck.  The compiler does not simply generate C, however, but C
  extended with calls to the ATerm library, a run-time library for
  term manipulation and storage (Sec.~\ref{sec:TermLib}). C cannot be
  changed, but the ATerm library can be adapted to prevent it from
  becoming an obstacle to further code improvement, should the need
  arise. We note, however, that the fact that the ATerm library
  interface is made available as an API to users outside the compiler
  makes it harder to adapt. 

  Although we feel these to be useful guidelines, they have to be
  applied with care. Their validity is not absolute, but depends on
  many details of the actual implementation under consideration.  The
  compiler for the lazy functional language Clean
  \cite{PlasmeijerVanEekelen:94,SmetsersEtAL:91}, for instance,
  generates native code via an abstract graph rewriting machine,
  contravening several of our guidelines.  Nevertheless, our benchmarks
  (Sec.~\ref{sec:BM}) show the Clean compiler and the \asfsdf\
  compiler to generate code with comparable performance.

\subsection{Organization of Term Storage} \label{sec:TERMS}

  \asfsdf\ applications may involve rewriting of large terms ($> 10^6$
  nodes).  Usually, this requires constructing and matching many
  intermediate results and the proper organization of term storage
  becomes critical to the run-time performance of the term datatype
  provided by the ATerm library and, as a consequence, to the run-time
  performance of the generated code as a whole.  Fortunately,
  intermediate results created during rewriting tend to have a lot of
  overlap.  This suggests use of a space saving scheme where terms are
  created only when they do not yet exist.  The various trade-offs
  involved in this choice are discussed in Sec.~\ref{sec:TermLib}.

\subsection{Separate Compilation} \label{sec:SEPCOMP}

      For large modularized language definitions, separate compilation is
      as important as it is for large modularized programs.  Fully
      separate compilation of \asfsdf\ modules is hard since the
      rewrite rules defining an \asfsdf\ function may be scattered
      over several modules and each \asfsdf\ function has to
      correspond to a single C function in the generated code for
      reasons of efficiency.  Fortunately, the number of modules
      contributing to the definition of an \asfsdf\ function is
      usually very small, so a useful approximation to separate
      compilation of \asfsdf\ modules can still be obtained.

\section{The \asfsdf\ Language} \label{sec:ASFSDF}

In addition to regular rewrite rules, \asfsdf\ features conditional
rewrite rules, associative (flat) lists, default rules, and simple
modularization.  In our discussion of these features we will emphasize
issues affecting their compilation.  A more detailed semantics
by example of \muasf, which helped to answer the questions that
emerged while the compiler was being written, is given by
\citeN{BergstraVandenBrand:00}.  For the use of \asfsdf\ (including
SDF) see \cite{DHK96}.

Since we do not go into the syntax definition component \sdf, we will
use a running example written in \muasf, the abstract syntax (prefix
notation only) of \asfsdf.  Consider the definition of a
simple type environment in Figure~\ref{fig:ENV}. The functions and
constants used in the rules are declared in the signature section,
with their argument positions (if any) indicated by
underscores. Although \asfsdf\ is a many-sorted formalism, the sorts
can be dispensed with after parsing and conversion to \muasf.  The
predefined list constructors {\tt list} (conversion to single element
list), {\tt conc} (associative list concatenation), and {\tt null}
(the empty list) need not be declared.

Symbols starting with a capital are variables. These are first-order,
i.e., they cannot have arguments, and need not be declared in the
signature. List variables are prefixed with a ``{\tt *}'' if they
can match the empty list or with a ``{\tt +}'' if they cannot.

The predefined symbols used in the rules are listed in
Table~\ref{table:SYMBOLS}. The example contains a single conditional
rule~{\tt [at-2]} with both a negative and a positive condition,
and a single default rule~{\tt [l-2]}. 
 
With an appropriate user-defined syntax, the \asfsdf\ version
of rule {\tt [at-1]} would get the more natural look
\begin{small}
\begin{verbatim}
[at-1] add (Id,Type1) to {(Id,Type2),Pair1*} = {(Id,Type1),Pair1*}; 
\end{verbatim}
\end{small} 
and similarly for the other rules.  In the following sections we
explain the various types of rules in more detail.

\begin{figure}[t]
\rule{12.5cm}{.5mm}
\begin{small}
\begin{verbatim}
module Type-environment
signature
  nil-type       {constructor};
  pair(_,_)      {constructor};       
  type-env(_)    {constructor};     
  lookup(_,_);     
  add-to(_,_,_)
rules
 
[l-1]  lookup(Id,type-env(conc(*Pair1,conc(pair(Id,Type),*Pair2))))
              = Type;

[l-2]  default: lookup(Id,Tenv) 
              = nil-type;

[at-1] add-to(Id,Type1,type-env(conc(pair(Id,Type2),*Pair1)))
              = type-env(conc(pair(Id,Type1),*Pair1));

[at-2] Id1 != Id2 &
       add-to(Id1,Type1,type-env(*Pair1)) == type-env(*Pair2)
         ==>
       add-to(Id1,Type1,type-env(conc(pair(Id2,Type2),*Pair1)))
              = type-env(conc(pair(Id2,Type2),*Pair2));

[at-3] add-to(Id,Type,type-env(null)) 
              = type-env(list(pair(Id,Type)))
\end{verbatim}
\end{small}
\rule{12.5cm}{.5mm}
\caption{Definition of a simple type environment in \muasf, the abstract
syntax (prefix notation only) version of \asfsdf\ produced by the
parsing phase.} \label{fig:ENV}
\end{figure}

\begin{table}
\begin{center}
\begin{tabular}{|l|l|} \hline
{\tt =}        & left-to-right rewrite \\ 
{\tt ==}       & equality in positive condition \\ 
{\tt !=}       & inequality in negative condition \\ 
{\tt \&}       & conjunction of conditions \\ 
{\tt ==>}      & implication \\
{\tt default:} & default rule flag \\
{\tt list}     & conversion to single element list \\
{\tt conc}     & associative list concatenation \\
{\tt null}     & empty list \\
\hline
\end{tabular}
\end{center}
\caption{The predefined symbols used in \muasf\ rewrite rules.} \label{table:SYMBOLS}
\end{table}

\subsection{Conditional Rewrite Rules} \label{sec:CONDRULES}

We assume throughout that the terms being rewritten are ground terms,
i.e., terms without variables.  A rule is applicable to a redex if its
left-hand side matches the redex and its conditions (if any) succeed
after substitution of the values found during matching.

Negative conditions succeed if both sides are syntactically different
after normalization. Otherwise they fail. They are not allowed to
contain variables not already occurring in the left-hand side of the
rule or in a preceding positive condition.  This means both sides of a
negative condition are ground terms at the time the condition is
evaluated.

Positive conditions succeed if both sides are syntactically equal
after normalization. Otherwise they fail.  One side of a positive
condition may contain one or more
new variables not already occurring in the left-hand side of
the rule or in a preceding positive condition.  This means one side of
a positive condition need not be a ground term at the time it is
evaluated, but may contain existentially quantified variables. Their
value is obtained by matching the side they occur in with the other
side after the latter has been normalized.  The side containing the
variables is not normalized before matching.

Variables occurring in the right-hand side of the rule must occur in
the left-hand side or in a positive condition, so the right-hand side
is a ground term at the time it is substituted for the redex.

Consider rule~{\tt [at-2]} in Fig.~\ref{fig:ENV} keeping the above
in mind. Its application proceeds as follows:
\begin{enumerate}

\item Find a redex matching the left-hand side of the rule (if any). 
      This yields values for the variables {\tt Id1},
      {\tt Type1}, {\tt Id2}, {\tt Type2}, and
      {\tt *Pair1}.
 
\item Evaluate the first condition. This amounts to a simple syntactic
      inequality check of the two identifiers picked up in step 1.  If
      the condition succeeds, evaluate the second one. Otherwise, the
      rule does not apply.

\item Evaluate the second condition. 
      This is a positive condition containing the new list variable
      {\tt *Pair2} in its right-hand side.  The value of
      {\tt *Pair2} is obtained by matching the right-hand side with
      the normalized left-hand side. Since {\tt *Pair2} is a list
      variable, this involves list matching, which is explained below.
      In this particular case, the match always succeeds.

\item Finally, replace the redex with the right-hand side of the rule
      after substituting the values of {\tt Id2} and {\tt Type2}
      found in step 1 and the value of {\tt *Pair2} found in step 3.

\end{enumerate}

\subsection{Lists} \label{sec:LISTS}

\asfsdf\ lists are associative (flat) and list matching is the same
as string matching.  Unlike a term pattern, a list pattern may match a
redex in more than one way. This may lead to backtracking within the
scope of the rule containing the list pattern in the following two
closely related cases:
\begin{itemize}

\item A rewrite rule containing a list pattern in its left-hand side
      might use conditions to select an appropriate match from the
      various possibilities.

\item A rewrite rule containing a list pattern with new variables
      in a positive condition (Sec.~\ref{sec:CONDRULES}) might use
      additional conditions to select an appropriate match from the
      various possibilities.

\end{itemize}

List matching may be used to avoid the explicit traversal of
structures.  Rule {\tt [l-1]} in Fig.~\ref{fig:ENV} illustrates this.
It does not traverse the type environment explicitly, but picks an
occurrence (if any) of the identifier it is looking for using two list
variables {\tt *Pair1} and {\tt *Pair2} to match its context.
The actual traversal code is generated by the compiler.  In general,
however, there is a price to be paid. While term matching is linear,
string matching is NP-complete \cite{BenanavEtAl:85}. Hence, list
matching is NP-complete as well. It remains an important source of
inefficiency in the execution of \asfsdf\ definitions \cite{Vinju99}.

\subsection{Default Rules} \label{sec:DEFAULTRULES}

A default rule has lower priority than ordinary rules in the sense
that it can be applicable to a redex only if all ordinary rules are
exhausted.  In Fig.~\ref{fig:ENV}, {\tt lookup} uses default
rule {\tt [l-2]} to return {\tt nil-type} if rule {\tt [l-1]}
fails to find the identifier it is looking for.

\subsection{Constructors}

A (free) constructor is a function that does not occur at the
outermost position in the left-hand side of a rewrite rule.  
A term consisting solely of constructors is in normal form.
In \asfsdf\ the rules defining a function may be scattered over many
modules, so this is a global property.  The {\tt constructor}
attribute supplies this information locally in a module, thus
improving readability and facilitating separate compilation of
modules. In Fig.~\ref{fig:ENV}, the functions {\tt nil-type}, {\tt
pair}, and {\tt type-env} are declared as constructors.  As mentioned
before, the built-in list constructors {\tt list}, {\tt
conc},\footnote{The associativity of {\tt conc} is taken care of by
list matching, otherwise it is a free constructor.} and {\tt null}
need not be declared.  Omitting constructor attributes is not a fatal
error, but may result in less readable \asfsdf\ definitions as well as
less efficient code. Some of the compiler optimizations depend
on constructor attributes being present in the \asfsdf\ source.

\subsection{Modules}

\asfsdf's only module operation is {\tt import}.\footnote{
The parameterization and renaming operations of \asf\ \cite{BHK89} are
not available in the current implementation of \asfsdf.}  
As mentioned in Sec.~\ref{sec:SEPCOMP}, separate
compilation of modules is an important design issue.

\subsection{Rewriting Strategies} \label{sec:STRAT}

\asfsdf\ is a strict language based on innermost rewriting (call-by-value).
With few exceptions, practical experience with \asfsdf\ over the past
ten years has shown innermost rewriting to be a good choice for
several reasons:
\begin{itemize}

\item Most users are familiar with call-by-value from C and
      other imperative languages.

\item It is consistent with the semantics of \asfsdf's default rules
      (Sec.~\ref{sec:DEFAULTRULES}).

\item Its behavior is more predictable than that of other strategies,
      an important consideration when rewrite systems become large.

\item No strictness annotations need to be added by the user to improve
      the quality of the code generated by the compiler. This is an
      advantage in view of the fact that ``inserting these strictness
      annotations correctly can be a fine art''
      \cite[p.~651]{HartelEtAl:96}.

\item It facilitates compilation to and interfacing with C and 
      other imperative languages. In particular, it allows \asfsdf\
      functions to be mapped directly to C functions and intermediate
      results produced during term rewriting to be stored in an
      efficient way (Sec.~\ref{sec:TermLib}).

\end{itemize}

We also encountered cases (conditionals, for instance) where innermost
rewriting proved unsatisfactory.  In such cases, rewriting of specific
function arguments can be delayed by annotating them with the
{\tt delay} attribute. See \cite{BergstraVandenBrand:00} for details.

\section{Preprocessing} \label{sec:PREPROC}

\begin{figure}
\rule{12.5cm}{.5mm}
\small

\begin{picture}(400,450)(-150,-430)
% last tuple = position

\thicklines

\put(0,0)
  {\begin{picture}(200,200)
      \put(0,0){\oval(70,30)}
      \put(0,0){\makebox(0,0){\asfsdf}}
      \put(0,-15){\vector(0,-1){30}}
      \put(10,-30){Parsing}
  \end{picture}
  }

\put(0,-60)
  {\begin{picture}(200,200)
      \put(0,0){\oval(70,30)}
      \put(0,0){\makebox(0,0){\muasf}}
      \put(0,-15){\vector(0,-1){140}}
      \put(10,-149){\shortstack[l]{
                    Preprocessing (Sec. \ref{sec:PREPROC}):\\
                     $\bigcirc$ Collection of rules per function\\
                     $\bigcirc$ Linearization of left-hand sides\\
                     $\bigcirc$ Introduction of assignments in conditions\\
                     $\bigcirc$ Elimination of constructor arguments from\\
                     \rule[-.1cm]{0cm}{.3cm} \hspace{.6cm} left-hand sides\\
                     $\bigcirc$ Simplification of patterns in assignment\\
                     \rule[-.1cm]{0cm}{.3cm} \hspace{.6cm} conditions\\
                     $\bigcirc$ Simplification of list patterns\\
                     $\bigcirc$ Combination of rules with identical\\
                     \rule[-.1cm]{0cm}{.3cm} \hspace{.6cm} conditions\\
                     $\bigcirc$ Introduction of \texttt{else} cases
                    }
                   }      
  \end{picture}
  }

\put(0,-230)
  {\begin{picture}(200,200)
      \put(0,0){\oval(70,30)}
      \put(0,0){\makebox(0,0){\muasfplus}}
      \put(0,-15){\vector(0,-1){60}}
      \put(10,-62){\shortstack[l]{
                    Code generation (Sec. \ref{sec:CODEGEN}):\\
                     $\bigcirc$ Term matching automata\\
                     $\bigcirc$ List matching code\\
                     $\bigcirc$ Memoization
                     }
                  }         
  \end{picture}
  }

\put(0,-320)
  {\begin{picture}(200,200)
      \put(0,0)  {\oval(150,30)}
      \put(0,0)  {\makebox(0,0){C $+$ ATerm Library primitives}}
      \put(0,-15){\vector(0,-1){60}}
      \put(10,-55){\shortstack[l]{
                    Postprocessing (Sec. \ref{sec:POSTPROC}):\\
                     $\bigcirc$ Tail recursion elimination\\
                     $\bigcirc$ Constant caching
                    }
                  }
  \end{picture}
  }

\put(0,-410)
  {\begin{picture}(200,200)
      \put(0,0){\oval(150,30)}
      \put(0,0){\makebox(0,0){C $+$ ATerm Library primitives}}
  \end{picture}
  }
\end{picture}
\normalsize
\rule{12.5cm}{.5mm}
\caption{Layout of the \asfsdf\ compiler. This is a refinement of Figure~\ref{fig:CompLayoutRough}.} \label{fig:CompLayout}
\end{figure}
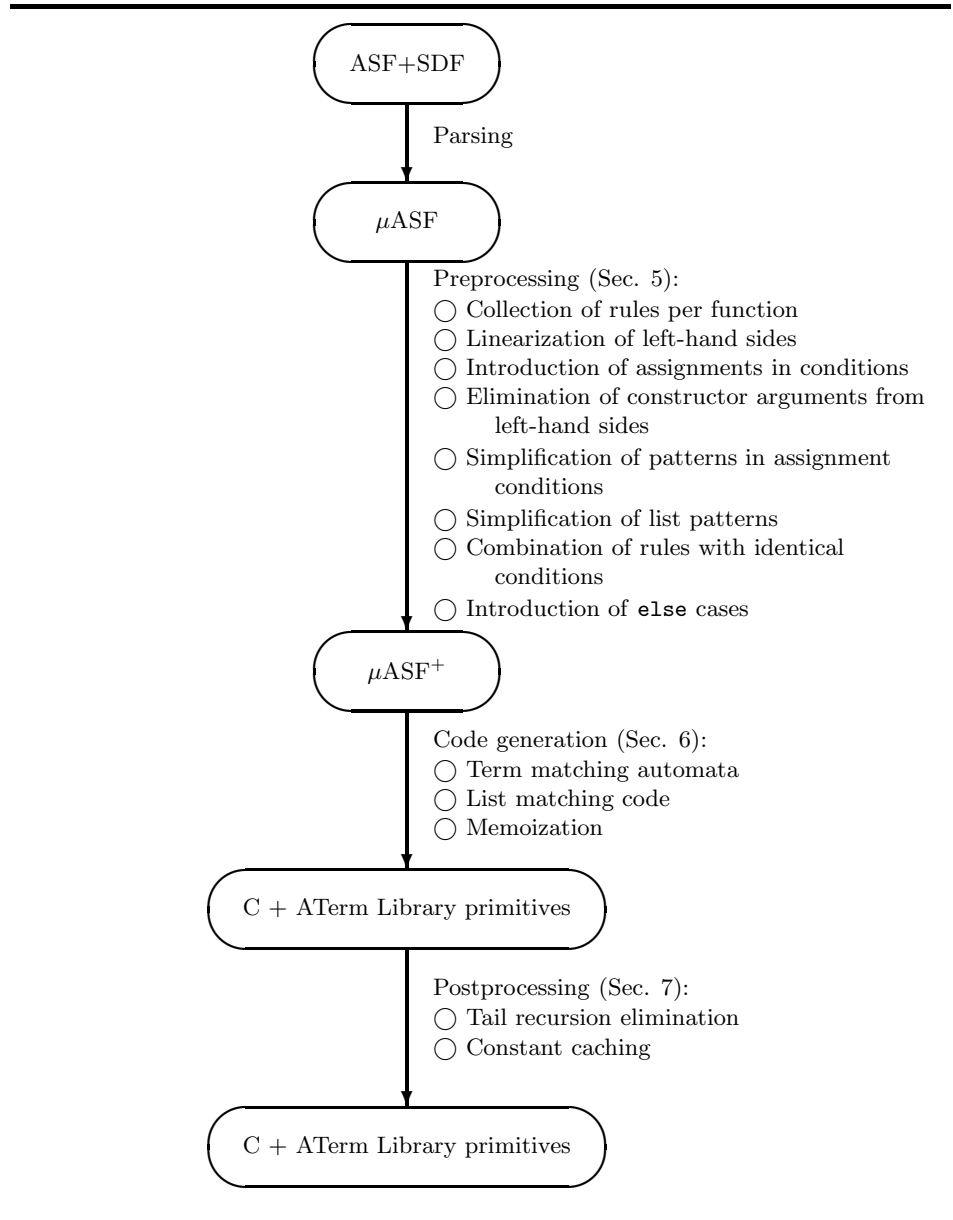

Figure~\ref{fig:CompLayout} is a refinement of
Figure~\ref{fig:CompLayoutRough} showing the preprocessing steps as
well as other actions performed in later phases of the compiler.  The
output language of the preprocessing phase is \muasfplus, which is
\muasf\ with the additional constructs shown in Table~\ref{table:ADDSYMBOLS}.
Their purpose will become clear later on when the preprocessing
(Sec.~\ref{sec:PREPROC}) and code generation (Sec.~\ref{sec:CODEGEN})
are discussed. Some of them, like nested rules, the {\tt
else}-construct, and the assignment, might very well be added to
\asfsdf\ itself, but this remains to be done.

\begin{table}
\begin{center}
\begin{tabular}{|l|l|} \hline
{\tt :=}           & assignment \\
{\tt \{ \}}        & nesting of rules \\
{\tt else}         & alternative \\
{\tt list\_head}   & first element of list \\
{\tt list\_tail}   & tail of list \\
{\tt list\_last}   & last element of list \\
{\tt list\_prefix} & prefix of list \\
{\tt not\_empty\_list} & list-not-empty predicate \\
{\tt t, f}         & true, false\\
\hline
\end{tabular}
\end{center}
\caption{Additional predefined symbols of \muasfplus.} \label{table:ADDSYMBOLS}
\end{table}
  
We now discuss the various preprocessing steps in more detail.  As
noted in Sec.~\ref{sec:PITFALLS}, they have to be chosen judiciously
to prevent them from becoming counterproductive, especially if their
purpose is to reduce the number of different constructs that have to
be handled by the code generator.  Each step has to preserve the
innermost rewriting strategy\footnote{Function arguments annotated
with the {\tt delay} attribute (Sec.~\ref{sec:STRAT}) have to be
taken into account as well, but will be ignored in this article for
the sake of readability.} as well as the backtracking behavior of list
matching.

\subsection{Collection of Rules per Function} \label{sec:collect}

As mentioned in Sec.~\ref{sec:SEPCOMP}, fully separate compilation of
\asfsdf\ modules is hampered by the fact that the rewrite rules for a
function can be scattered over several modules.  Given a top module
for which an executable has to be generated, the preprocessing phase
starts by traversing the top module and all modules directly and
indirectly imported by it, collecting the rewrite rules for each
function declared in its signature, i.e., the rules whose left-hand
side has the function as its outermost symbol.  The rules collected
for each function together with the corresponding function declaration
from the signature are made into a new \muasf\ module.\footnote{For
reasons of efficiency, constructor functions (which can never occur at
the outermost position of a left-hand side) are not made into separate
modules. Instead, the constructors defined in a module are kept
together and made into a single new module.}  When a rewrite rule is
changed, only the module containing the function actually affected is
recompiled.  This yields a useful approximation to separate
compilation because the number of modules involved is usually limited
($< 100$) and the number of modules contributing to the definition of
a function is usually very small. Still, the full specification
has to be scanned for the rare cases a function is not completely
defined in a single module, and a function attribute ruling this out
would be a useful addition to \asfsdf.

\subsection{Linearization of Left-Hand Sides} \label{sec:LIN}

A rewrite rule is \emph{non-linear} if its left-hand side contains
more than one occurrence of the same variable. Different occurrences
of the same variable have to obtain the same value during matching, so
non-linearity amounts to an implicit equality check.  Non-linearities
are eliminated by adding appropriate positive conditions.  Innermost
rewriting guarantees that these conditions do not cause spurious
rewrite steps not done by the original non-linear
match.\footnote{Non-linearities involving function arguments annotated
with the {\tt delay} attribute are not allowed.}

For example, rules {\tt [l-1]} and {\tt [at-1]} in
Fig.~\ref{fig:ENV} are non-linear since variable {\tt Id} occurs
twice in their left-hand side. Rule {\tt [at-1]} would be
transformed into
\begin{small}
\begin{verbatim}
[at-1'] Id == Id1    
          ==>
        add-to(Id,Type1,type-env(conc(pair(Id1,Type2),*Pair1)))
              = type-env(conc(pair(Id,Type1),*Pair1))
\end{verbatim}
\end{small}
with new variable {\tt Id1} not already occurring in the original
rule, and similarly for {\tt [l-1]}.

Linearization has pros and cons. On the one hand, it simplifies the
matching automaton and enables further transformations, especially the
introduction of {\tt else}s if there is a corresponding rule with a
negative condition as is often the case (see below).  The condition is
implemented very efficiently as a pointer equality check as will be
explained in Sec.~\ref{sec:TermLib}.  On the other hand, in rare cases it may also cause
inefficiencies.  Consider, for instance, a rule {\tt f(X,X,lp) =
$\ldots$} with complicated list pattern {\tt lp}. A straightforward
implementation would first check the equality of the values obtained
for the first two arguments of {\tt f} before proceeding with the
matching of {\tt lp}.  A straightforward implementation of the
transformed rule\\

{\tt X == X1 ==> f(X,X1,lp) = $\ldots$}\\

\noindent as currently generated by the
compiler postpones the equality check and does a full match of {\tt
f(X,X1,lp)} first.  This is inefficient if the full match
succeeds with unequal values for {\tt X} and {\tt X1}.

\subsection{Introduction of Assignments in Conditions}

As explained in Sec.~\ref{sec:CONDRULES}, one side of a positive
condition may contain variables that are uninstantiated at the time
the condition is evaluated. Their value is obtained by matching the
side they occur in with the other side after the latter has been
normalized. The side containing the uninstantiated variables is not
normalized before matching.  To flag this case to the code generation
phase, the \muasf\ equality is replaced by the \muasfplus\ assignment.
If necessary, the left- and right-hand side of the original condition
are interchanged.

Rule {\tt [at-2]} in Fig.~\ref{fig:ENV} is of this
kind since its second condition contains the new list variable
{\tt *Pair2}.  It would be transformed into
\begin{small}
\begin{verbatim}
[at-2'] Id1 != Id2 &
        type-env(*Pair2) := add-to(Id1,Type1,type-env(*Pair1))
          ==>
        add-to(Id1,Type1,type-env(conc(pair(Id2,Type2),*Pair1)))
              = type-env(conc(pair(Id2,Type2),*Pair2)).
\end{verbatim}
\end{small}

\subsection{Elimination of Constructor Arguments from Left-Hand Sides} \label{sec:ELIMARG}

Complex arguments consisting solely of constructors are eliminated
from left-hand sides of rules and moved to assignment conditions. Let
{\tt f($\ldots$,ct,$\ldots$) = $\ldots$} be such a rule with complex
constructor term {\tt ct}. It is transformed to\\

{\tt X := ct ==> f($\ldots$,X,$\ldots$) = $\ldots$}.\\

\noindent This transformation simplifies the matching automaton by
replacing the matching of {\tt ct} by a simple pointer equality check
(this will become clear later). Since the value of {\tt X} is not
evaluated and {\tt ct} is already in normal form, it does not
introduce spurious rewrite steps not done by the original rule.

\subsection{Simplification of Patterns in Assignment Conditions}

If not already in the right form, assignment conditions will be broken
up into several new assignment conditions in such a way that the
patterns making up their left-hand sides consist of a single variable,
a single constant, or a single function symbol with only variables as
arguments.  This transformation has no effect on the performance or
even the structure of the corresponding matching automaton, but makes
its generation easier.

Rule {\tt [at-2']} has an assignment condition whose left-hand side
is already in the right form, so we give another example. The rule\\

{\tt g(h(a),Z) := k(X) ==> f(X,Y) = $\ldots$}\\

\noindent is transformed into\\

{\tt g(H,Z) := k(X) \& h(A) := H \& a := A ==> f(X,Y) = $\ldots$}.\\

\noindent In both the original and the transformed version, the
instantiated right-hand side {\tt k(X)} is normalized before the
assignment is evaluated by matching with its left-hand side.  Hence,
the values obtained for {\tt H} and {\tt A} (if any) by matching 
must themselves be normal forms, and the second and third assignment
cannot introduce spurious rewrite steps not done by the original
assignment.

\subsection{Simplification of List Patterns} \label{sec:SIMPLIST}

To simplify the generation of list matching code, list patterns in the
left-hand side of a rule or an assignment are brought in a standard
form containing, apart from the list constructors {\tt list} and
{\tt conc}, only variables and constants. Other more complicated
subpatterns are replaced by new variables that are evaluated in new
assignment conditions. This transformation preserves the backtracking
behavior of list matching, but may occasionally cause inefficiencies 
similar to those that may be caused by linearization (Sec.~\ref{sec:LIN}).

Rule {\tt [at-1']}, for example, will be transformed into
\begin{small}
\begin{verbatim}
[at-1''] pair(Id1,Type2) := P &
         Id == Id1
           ==>
         add-to(Id,Type1,type-env(conc(P,*Pair1)))
               = type-env(conc(pair(Id,Type1),*Pair1))
\end{verbatim}
\end{small}
and similarly for {\tt [at-2']} and {\tt [l-1]}.

List matching may cause backtracking, but list patterns containing
only a single list variable or no list variables at all never do. In
such cases, list matching can be eliminated using the
\muasfplus\ list functions in Table~\ref{table:ADDSYMBOLS}.
For example, {\tt [at-1'']} is transformed into
\begin{small}
\begin{verbatim}
[at-1'''] t := non_empty_list(*Pair) &
          P := list_head(*Pair) &
          *Pair1 := list_tail(*Pair) &
          pair(Id1,Type2) := P &
          Id == Id1
            ==>
          add-to(Id,Type1,type-env(*Pair))
                = type-env(conc(pair(Id,Type1),*Pair1)),
\end{verbatim}
\end{small}
where {\tt t} is the boolean value {\tt true}
(Table~\ref{table:ADDSYMBOLS}), and similarly for {\tt [at-2'']}.

\subsection{Combination of Rules with Identical Conditions}

Rules {\tt [at-1''']} and {\tt [at-2''']} resulting from the
previous step have their left-hand side and first four conditions in
common (up to renaming of variables). By factoring out the common
elements after a suitable renaming of variables, they can be combined
into the single nested rule
\begin{small}
\begin{verbatim}
[at-1-2]  t := non_empty_list(*Pair) &
          P := list_head(*Pair) &
          *Pair1 := list_tail(*Pair) &
          pair(Id1,Type2) := P
            ==>
          add-to(Id,Type1,type-env(*Pair)) = 
          {
          Id == Id1
            ==>
          type-env(conc(pair(Id,Type1),*Pair1));

          Id != Id1 &
          type-env(*Pair2) := add-to(Id,Type1,type-env(*Pair1))
            ==>
          type-env(conc(pair(Id1,Type2),*Pair2))
          },
\end{verbatim}
\end{small}
where the accolades are in \muasfplus. The depth of nesting produced
in this way may be arbitrarily large.

\subsection{Introduction of {\tt else} Cases} \label{sec:ELSE}

\muasfplus\ provides an {\tt else} construct which is
used to combine pairs of conditional rewrite rules with identical
left-hand sides (up to renaming of variables) and complementary
conditions. Introducing it in the result of the previous step yields
\begin{small}
\begin{verbatim}
[at-1-2'] t := non_empty_list(*Pair) &
          P := list_head(*Pair) &
          *Pair1 := list_tail(*Pair) &
          pair(Id1,Type2) := P
            ==>
          add-to(Id,Type1,type-env(*Pair)) = 
          {
          Id == Id1
            ==>
          type-env(conc(pair(Id,Type1),*Pair1))
            else
          type-env(*Pair2) := add-to(Id,Type1,type-env(*Pair1))
            ==>
          type-env(conc(pair(Id1,Type2),*Pair2))
          }.
\end{verbatim}
\end{small}

\section{Code Generation} \label{sec:CODEGEN}

\subsection{The ATerm Library} \label{sec:TermLib}

\subsubsection{Introduction}

The compiler generates C extended with calls to the ATerm library, a
run-time library for term manipulation and storage.  In this section
we discuss the ATerm library from the perspective of the compiler.
For a broader viewpoint and further applications see
\cite{VandenBrandEtAl:99,VandenBrandEtAl:00}.

Selected ATerm library functions are listed in
Table~\ref{table:ATermfunctions}.  Many of them correspond directly to
predefined symbols of \muasf\ (Table~\ref{table:SYMBOLS}) and
\muasfplus\ (Table~\ref{table:ADDSYMBOLS}). 
Examples of actual code using them is given in Sec.~\ref{sec:MATCHING}
and Sec.~\ref{sec:CONDSRHS}.

\begin{table}
\begin{center}
\begin{tabular}{|l|l|} \hline
{\tt term\_equal(t1,t2)}   & Check if terms {\tt t1} and {\tt t2} are equal \\
{\tt make\_list(t)}        & Create list with {\tt t} as single element \\
{\tt conc(l1,l2)}          & Concatenate lists {\tt l1} and {\tt l2} \\ 
{\tt null()}               & Create empty list \\
{\tt list\_head(l)}        & Get head of list {\tt l} \\
{\tt list\_tail(l)}        & Get tail of list {\tt l} \\
{\tt list\_last(l)}         & Get last element of list {\tt l} \\
{\tt list\_prefix(l)}       & Get prefix of list {\tt l} \\
{\tt not\_empty\_list(l)}  & Check if list {\tt l} is empty \\
{\tt is\_single\_element(l)} & Check if list {\tt l} has a single element \\
{\tt slice(p1, p2)}        & Take slice of list starting at pointer {\tt p1} and \\
                           & ending at {\tt p2} \\   
{\tt check\_sym(t,s)}   & Check if term {\tt t} has outermost symbol {\tt s} \\
{\tt arg\_i(t)}            & Get {\tt i}-th argument  \\
{\tt make\_nfi(s,t0,...,ti-1)} & Construct normal form with outermost \\
                  & symbol {\tt s} and arguments {\tt t0},$\ldots$,{\tt ti-1} \\
\hline
\end{tabular}
\end{center}
\caption{Selected ATerm library functions.} \label{table:ATermfunctions}
\end{table}

\subsubsection{Term Storage} \label{sec:TermStorage} 

The decision to store terms uniquely, which was briefly discussed in
Sec.~\ref{sec:TERMS}, is a major factor in the good run-time
performance of the code generated by the compiler.  If a term to be
constructed during rewriting already exists, it is reused, thus
guaranteeing maximal sharing.  This strategy exploits the redundancy
typically present in the terms built during rewriting.  The sharing is
transparent, so the compiler does not have to take precautions
during code generation. 

Maximal sharing of terms can only be maintained if the term construction
functions {\tt make\_nf0}, {\tt make\_nf1}, $\ldots$
(Table~\ref{table:ATermfunctions}) check whether the term to be
constructed already exists.  This implies a search through all existing
terms which must be very fast in order not to impose an unacceptable
penalty on term construction.  Using a hash function depending on the
internal code of the function symbol and the addresses of its
arguments, {\tt make\_nf}$i$ can quickly search for a function
application before constructing it.  Hence, apart from the space overhead
caused by the initial allocation of a hash table of sufficient
size,\footnote{Hash table overflow is not fatal, but causes allocation
of a larger table followed by rehashing.} the modest (but not
negligible) time overhead at term construction time is one hash table
lookup.

We get two returns on this investment.  First, the amount of space
gained by sharing terms is usually much larger than the space used by
the hash table. This is useful in itself, but it also yields a
substantial reduction in (real-time) execution time.  Second, {\tt
term\_equal}, the equality check on terms, only has to check for
pointer equality rather than structural equality.  The compiler
generates calls to {\tt term\_equal} in the pattern matching and
condition evaluation code.  For the same reason, this storage scheme
combines very well with memoization (Sec.~\ref{sec:Memoization}).

\subsubsection{Shared Terms vs. Destructive Updates}

Shared terms cannot be modified without causing unpredictable
side-effects, the more so since the ATerm library is not only used by
compiler generated code but also by other components of
the \asfsdf\ Meta-Environment.  Destructive updates would therefore cause
unwanted side-effects throughout the system.

During rewriting by compiler generated code the immutability
of terms causes no efficiency problems since they are created in a
non-destructive way as a consequence of the innermost reduction
strategy. Normal forms are constructed bottom-up and there is no need
to perform destructive updates on a term once it has been
constructed. Also, during normalization the input term itself is not
modified but the normal form is constructed separately.  Modification
of the input term would result in graph rewriting instead of (innermost)
term rewriting.

List operations like concatenation and slicing may become expensive,
however, if they cannot simply modify one of their arguments.  List
concatenation, for instance, can only be performed using ATerm library
primitives by taking the second list, successively prepending the
elements of the first list to it, and returning the new list as a
result.

The idea of subterm sharing is known in the Lisp community as
\emph{hash-consing} \cite{Alan:78}. Its success has been limited
by the existence of the Lisp functions {\tt rplaca} and
{\tt rplacd}, which modify a list destructively.
HLisp (Hash Lisp) is a Lisp dialect supporting hash-consing at the
language level \cite{TerashimaKanada:90}.  It has two kinds of list
structures: ``monocopy'' lists with maximal sharing and ``multicopy''
lists without maximal sharing.  Before a destructive change is made to
a monocopy list, it has to be converted to a multicopy list.

\asfsdf\ does not have functions like {\tt rplaca} and
{\tt rplacd}, and the ATerm library only supports the
equivalent of HLisp monocopy lists. Although the availability of
destructive updates would make the code for some list operations more
efficient, such cases are relatively rare. This explains why the
technique of subterm sharing can be applied more successfully in
\asfsdf\ than in Lisp.

Our positive experience with hash-consing in \asfsdf\ refutes the
theoretical arguments against its potential usefulness in the
equational programming language Epic mentioned by
\citeN[p.~701]{FokkinkEtAl:98}.  Also, while our experience seems to
be at variance with observations made by
\citeN{AppelGoncalves:93} in the context of SML, where sharing resulted
in only slightly better execution speed and marginal space savings,
both sharing schemes are actually rather different.  In our scheme,
terms are shared immediately at the time they are created, whereas
\citeANP{AppelGoncalves:93} delay the sharing of subterms until the
next garbage collection.  This minimizes the overhead at term
construction time, but at the same time sacrifices the benefits (space savings
and a fast equality test) of sharing terms that have not yet survived
a garbage collection.  The different usage patterns of terms in SML
and \asfsdf\ may also contribute to these seemingly contradictory
observations.

\subsubsection{Garbage Collection}

During rewriting, a large number of intermediate results is created,
most of which will not be part of the end result and have to be
reclaimed. There are basically three realistic alternatives for
this. We will discuss their advantages and disadvantages in relation
to the ATerm library. For an in-depth discussion of garbage collection
in general and these three alternatives in particular, we refer the
reader to \citeN{JonesLins:96}.

Since ATerms do not contain cycles, reference counting is an obvious
alternative to consider. Two problems make it unattractive, however.
First, there is no portable and efficient way in C to detect when
local variables are no longer in use.  Second, the memory overhead of
reference counting is large. Most ATerms can be stored in a few
machine words, and it would be a waste of memory to add another word
solely for the purpose of reference counting.

The other two alternatives are mark-compact and mark-sweep garbage
collection. The choice of C as an implementation language is not
compatible with mark-compact garbage collection since there is no
portable and at the same time reliable way in C to find all local
variables on the stack without help from the programmer.  This means
pointers to ATerms on the stack cannot be made to point to the new
location of the corresponding terms after compactification. The usual
solution is to ``freeze'' all objects that might be referenced from
the stack, and only relocate objects that are not.  Not being able to
move all terms negates many of the advantages of mark-compact garbage
collection such as decreased fragmentation and fast allocation.

The best alternative turns out to be mark-sweep garbage collection.
It can be implemented efficiently in C, both in time and space, and
with little or no support from the programmer \cite{Boehm:93}.  We
implemented this garbage collector from scratch, with many of the
underlying ideas taken directly from \citeANP{Boehm:93}'s garbage
collector, but tailored to the special characteristics of ATerms both
to obtain better control over the garbage collection process as well
as for reasons of efficiency.

Starting with the former, ATerms are always referenced from a hash
table, even if they are no longer in use. Hence, the garbage collector
should not scan this table for references.  We also need enough
control to remove an ATerm from the hash table when it is freed,
otherwise the table would quickly fill up with unused term references.

As for efficiency, experience shows that typically very few ATerms are
referenced from static variables or from generic datastructures on the
heap.  By providing a mechanism ({\tt ATprotect}) to enable the user
of the ATerm library to register references to ATerms that are not
local (auto) variables, we are able to completely eliminate the
expensive scan of the static data area and the heap.

We also have the advantage that almost all ATerms can be stored using
only a few words of memory. This makes it convenient to base the
algorithm used on only a small number of block sizes compared to a
generic garbage collector that cannot make any assumptions about the
sizes of the memory chunks that will be requested at run-time.

\subsection{Matching} \label{sec:MATCHING}

\subsubsection{Term Matching} \label{sec:termmatching}

After collecting the rules making up a function definition
(Sec.~\ref{sec:collect}), the compiler transforms their left-hand
sides into a deterministic finite automaton that controls the matching
of the function call at run-time, an approach originally due to
\citeN{HoffmannODonnell82}.  For reasons of separate compilation, each
generated C function has its own local matching automaton, unlike, for
instance, the compiler for the Elan rewriting logic language
\cite{MoreauKirchner:98}, which generates
a single large matching automaton.

The semantics of \asfsdf\ does not prescribe a particular way to
resolve ambiguous matches, i.e., more than a single left-hand side
matching the same innermost redex, so the compiler is free to choose a
suitable disambiguation strategy. To obtain a deterministic matching
automaton it uses the specificity order defined in
\cite[Definition~2.2.1]{FokkinkEtAl:98}.  Rewrite rules with more
specific left-hand sides take precedence over rules whose left-hand
sides are more general. Default rules correspond to ``otherwise''
cases in the automaton.

In the generated C code the matching automata are often hard to
distinguish from the conditions of conditional rules, especially since
the latter may have been generated in the preprocessing phase by the
compiler itself to linearize or simplify left-hand sides.

The matching automata generated by the compiler are not necessarily
optimal. We decided to keep the compiler simple, and take the
suboptimal code for granted, especially since it usually does not make
much difference.  Consider the following two rules
\begin{verbatim}
  f(a,b,c) = g(a)
  f(X,b,d) = g(X),
\end{verbatim}
where {\tt a}, {\tt b}, {\tt c}, {\tt d} are constants, and {\tt X} is
a variable.  The compiler currently generates the following
code in this case:
\begin{small}
\begin{verbatim}
  ATerm f(ATerm arg0, ATerm arg1, ATerm arg2) {
    if term_equal(arg0,a) { 
       if term_equal(arg1,b) { 
          if term_equal(arg2,c) {
             return g(a);
          }
       }
    }
    if term_equal(arg1,b) {
       if term_equal(arg2,d) { 
          return g(arg0);
       }
    }
    return make_nf3(fsym, arg0, arg1, arg2)
  },
\end{verbatim}
\end{small}
where {\tt fsym} is a constant corresponding to the function name {\tt f}.
The generated matching automaton is straightforward. It checks the
arguments of each left-hand side from left to right using the ATerm
library function {\tt term\_equal}, which does a simple pointer
equality check (Sec.~\ref{sec:TermStorage}).  If neither left-hand
side matches, the appropriate normal form is constructed by ATerm
library function {\tt make\_nf3} (Table~\ref{table:ATermfunctions}).

Slightly better code could be obtained by dropping the left-to-right
bias of the generated automaton\footnote{\citeN{NedjahEtAl:97} discuss
optimization of the matching automaton under a left-to-right
constraint.} and checking {\tt arg1} rather than {\tt arg0} first:
\begin{small}
\begin{verbatim}
  ATerm f(ATerm arg0, ATerm arg1, ATerm arg2) {
    if term_equal(arg1,b) {
       if term_equal(arg0,a) {
          if term_equal(arg2,c) { 
             return g(a);
          }
       }
       else if term_equal(arg2,d) {
            return g(arg0);
       }
    }
    return make_nf3(fsym, arg0, arg1, arg2)
  }.
\end{verbatim}
\end{small}

\subsubsection{List Matching}  \label{sec:listmatching}

\begin{figure}[t]
\rule{12.5cm}{.5mm}
\begin{small}
\begin{verbatim}
ATerm set(ATerm arg0) {
  ATerm tmp_0 = arg0;                    /* cursor in argument list */
  ATerm tmp_1[2];                        /* *Id0 (begin and end cursor) */
  tmp_1[0] = tmp_0;
  tmp_1[1] = tmp_0;
  while(not_empty_list(tmp_0)) {
    ATerm tmp_2[2];                      /* *Id1 (begin and end cursor) */
    ATerm tmp_3 = list_head(tmp_0);      /* Id */
    tmp_0 = list_tail(tmp_0);      
    tmp_2[0] = tmp_0;
    tmp_2[1] = tmp_0;
    while(not_empty_list(tmp_0)) {
      ATerm tmp_4 = list_head(tmp_0);    /* Id' */
      tmp_0 = list_tail(tmp_0);
      if(term_equal(tmp_3, tmp_4)) {     /* Id = Id' */
        return set(conc(slice(tmp_1[0], tmp_1[1]),
                   conc(tmp_3, conc(slice(tmp_2[0],tmp_2[1]), tmp_0))));
      }
      tmp_2[1] = list_tail(tmp_2[1]);
      tmp_0 = tmp_2[1];
    }
    tmp_1[1] = list_tail(tmp_1[1]);
    tmp_0 = tmp_1[1];
  }
  return make_nf1(setsym,arg0);
}              
\end{verbatim}
\end{small}
\rule{12.5cm}{.5mm}
\caption{Code generated for rule {\tt [s-1']}.} \label{fig:SETCODE}
\end{figure}

As was pointed out in Sec.~\ref{sec:SIMPLIST}, a few simple cases of
list matching that do not need backtracking are transformed to
ordinary term matching in the preprocessing phase.  The other cases
are translated to nested while-loops. These handle the (limited form of)
backtracking that may be caused by condition failure (Sec.~\ref{sec:LISTS}).

Consider the \asfsdf\ rule
\begin{verbatim}
[s-1] {Id0*,Id,Id1*,Id,Id2*} = {Id0*,Id,Id1*,Id2*}, 
\end{verbatim}
which makes lists into sets by removing elements that occur more than once. 
Its \muasf\ representation would be
\begin{verbatim}
[s-1] set(conc(*Id0,conc(Id,conc(*Id1,conc(Id,*Id2))))) =
      set(conc(*Id0,conc(Id,conc(*Id1,*Id2)))),
\end{verbatim}
where {\tt set} is some prefix representation of the user-defined
accolade notation for sets used in the \asfsdf\ rule, and {\tt conc}
is the predefined associative list concatenation of \muasf.  Each
application of {\tt [s-1]} picks up the leftmost pair of elements
occurring more than once in variable {\tt Id} and keeps only a single
occurrence in its right-hand side. List variables {\tt *Id1}, {\tt
*Id2}, and {\tt *Id3}, each of which can match the empty list, are
used to pick up and transfer the context.
 
Since rule {\tt [s-1]} is nonlinear, it is first transformed to 
\begin{verbatim}
[s-1'] Id == Id' 
         ==>
       set(conc(*Id0,conc(Id,conc(*Id1,conc(Id',*Id2))))) =
       set(conc(*Id0,conc(Id,conc(*Id1,*Id2))))
\end{verbatim}
by the preprocessor. The C code generated for rule {\tt [s-1']} is
shown in Fig.~\ref{fig:SETCODE}. It consists of two nested
while-loops, which try successive values for the three list variables.
The various ATerm functions used in it are listed in
Table~\ref{table:ATermfunctions}. The condition is checked in
the body of the innermost loop.

Rule {\tt [s-1]} is applied as often as needed to reach a normal form
containing each element only once, but each application is independent
of the previous one, starting from the beginning of the set rather
than at the position where the previous application left off.  This
leaves room for further optimization, but its implementation in
sufficiently general form to be effective has turned out to be hard \cite{Vinju99}.

\subsection{Evaluation of Conditions and Right-Hand Sides} \label{sec:CONDSRHS}

\begin{figure}[t]
\rule{12.5cm}{.5mm}
\begin{small}
\begin{verbatim}
ATerm add_to(ATerm arg0, ATerm arg1, ATerm arg2) 
{
  ATerm tmp[6];
  if (check_sym(arg2, extfun1_sym)) {     /* arg2 = type-env(*Pair) */
    ATerm atmp20 = arg_0(arg2);
    if (not_empty_list(atmp20)) {         /* t := non_empty_list(*Pair) */
      tmp[0] = list_head(atmp20);         /* P := list_head(*Pair) */
      tmp[1] = list_tail(atmp20);         /* *Pair1 := list_tail(*Pair) */
      if (check_sym(tmp[0], extfun2_sym)) {     /* pair(Id1,Type2) := P */
        tmp[2] = arg_0(tmp[0]);           /* Id1 */
        tmp[3] = arg_1(tmp[0]);           /* Type2 */
        if (term_equal(arg0, tmp[2])) {   /* Id == Id1 */
          return (*extfun1)(conc((*extfun2)(arg0, arg1), tmp[1]));

        }
        else {
          tmp[4] = add_to(arg0, arg1, (*extfun1)(tmp[1]));
                          /* tmp[4] = add-to(Id,Type1,type-env(*Pair1)) */
          if (check_sym(tmp[4], extfun1_sym)) {  
                                            /* tmp[4] = type-env(*Pair) */
            tmp[5] = arg_0(tmp[4]);
            return (*extfun1)(conc((*extfun2)(tmp[2], tmp[3]), tmp[5]));
          }
        }
      }
    }
    else {
      return (*extfun1)(make_list((*extfun2)(arg0, arg1)));
    }
  }
  return make_nf3(extfun1_sym, arg0, arg1, arg2);
}  
\end{verbatim}
\end{small}
\rule{12.5cm}{.5mm}
\caption{Code generated for rule {\tt [at-1-2']}.} \label{fig:ELSECODE}
\end{figure}

The code generated for rule {\tt [at-1-2']} (Sec.~\ref{sec:ELSE}) is
shown in Fig.~\ref{fig:ELSECODE}. Before execution starts, {\tt
*extfun1} and {\tt *extfun2} are linked dynamically to, respectively,
C functions {\tt type\_env} and {\tt pair}.  The reasons for doing
this at run-time are explained in Sec.~\ref{sec:DL}.  As in the
previous example, the various ATerm functions used in the code are
listed in Table~\ref{table:ATermfunctions}. The \muasfplus\ {\tt else}
of the rule corresponds to the first {\tt else} in the C code.

\subsection{Memoization} \label{sec:Memoization}

To obtain faster code, the compiler can be instructed to memoize explicitly
given \asfsdf\ functions. The corresponding C functions get local hash
tables to store each set of arguments\footnote{Function arguments
annotated with the {\tt delay} attribute need not be in normal form
when stored in the memo table.} along with the corresponding result
(normal form) once it has been computed.  When called with a ``known''
set of arguments, the result is obtained from the memo table rather
than recomputed.  See also \citeN[Chapter~19]{FieldHarrison:88}.  

Maximal subterm sharing (hash-consing) as used in the ATerm library
(Sec.~\ref{sec:TermStorage}) combines very well with memoization.
Since memo tables tend to contain many similar terms (function calls),
memo table storage is effectively reduced by sharing. Furthermore, the
check whether a set of arguments is already in the memo table is a
simple equality check on the corresponding pointers.  There is
currently no hard limit on the size of a memo table, so the issue of
replacement of table entries does not (yet) arise.

Unfortunately, since its effects may be hard to predict, memoization
is something of a ``fine art'', not unlike adding strictness
annotations to lazy functional programs.  Memoization may easily
become counterproductive if the memoized functions are not called with
the same arguments sufficiently often, and finding the right subset of
functions to memoize may require considerable experimentation and
insight.

\subsection{Dynamic Linking of \asfsdf\ Function Identifiers} \label{sec:DL}

\begin{table}
\begin{center}
\begin{tabular}{|l|l|} \hline
{\tt register\_prod(prod, funptr, symbol)} & Add C function pointer {\tt funptr} and unique \\
                          & symbol {\tt symbol} generated for function with \\
                          & \asfsdf\ identifier {\tt prod} to symbol table \\ 
{\tt lookup\_func(prod})  & Get C function pointer for function with \\
                          & \asfsdf\ identifier {\tt prod} \\
{\tt lookup\_sym(prod)}   & Get symbol for function with \asfsdf\ \\
                          & identifier {\tt prod} \\
{\tt lookup\_prod(symbol)}   & Return \asfsdf\ identifier of symbol {\tt symbol} \\
\hline
\end{tabular}
\end{center}
\caption{ATerm library functions used for dynamic linking.} \label{table:DLfunctions}
\end{table}

Because of the user-defined syntax, an \asfsdf\ function identifier
corresponds to an \sdf\ grammar production (which is similar to a BNF
rule). Mapping such rules to C function identifiers directly is not
possible because of length and character set restrictions.  To
circumvent this problem, we adopted a dynamic linking approach for
function identifiers in addition to the usual static linking.

More specifically, for each C file {\tt M} the compiler maps \asfsdf\
function identifiers (productions) to C function identifiers whose
uniqueness is not guaranteed beyond the scope of {\tt M}. This does
not require global knowledge. The compiler also generates additional
functions {\tt register\_M} and {\tt lookup\_M} for each C file {\tt
M}.  These are executed before actual rewriting starts and perform the
dynamic linking on the basis of the \asfsdf\ function
identifiers.  For each function defined in {\tt M}, {\tt register\_M}
stores the \asfsdf\ identifier along with the corresponding
unique C function pointer supplied by the preceding static linkage
editing phase in a symbol table using ATerm function {\tt
register\_prod} (Table~\ref{table:DLfunctions}).  For each external
function called from {\tt M}, {\tt lookup\_M} then obtains a pointer
from the symbol table on the basis of the \asfsdf\ identifier
using ATerm library function {\tt lookup\_func}.

\section{Postprocessing} \label{sec:POSTPROC}

The quality of the generated C code is further improved by tail
recursion elimination and constant caching.  Not all C compilers are
capable of tail recursion elimination, and no compiler known to us can
do it if it has to produce code with symbolic debugging information, so
the \asfsdf\ compiler takes care of this itself.  In principle, this
optimization could also be done by the preprocessor if a
while-construct were added to \muasfplus.

Constant caching is a restricted form of memoization.  Unlike the
latter, it is performed fully automatically on ground terms occurring
in right-hand sides of rules or in conditions.  These may be evaluated
more than once during the evaluation of a term, but since their normal
form is the same each time (no side-effects), they are recognized and
transformed into constants.  The first time a constant is encountered
during evaluation, the associated ground term is normalized and the
result is assigned to the constant.  In this way, the constant acts as
a cache for the normal form.

There are good reasons to prefer this hybrid compile-time/run-time
approach to a compile-time only approach:
\begin{itemize}
\item The compiler would have to normalize the ground terms in
      question. Although a suitable \muasf\ interpreter that can be
      called by the compiler exists, such normalizations potentially
      require the full definition to be available.  This is in
      conflict with the requirement of separate compilation.

\item The resulting normal forms may be quite big, causing
      an enormous increase in code size.
\end{itemize}

\section{Benchmarking} \label{sec:BM}

\begin{table}[t]
\begin{minipage}[c]{12cm}
\begin{tabular}{|l|l|l|} \hline
                  & Type of language and             &  \\       
Language          & semantic characteristics         & Compiled to       \\ 
\hline \hline
\asfsdf\          & Language definition formalism   & C \\
                  & $\bullet$ First-order           & \\
                  & $\bullet$ Strict                 & \\
                  & $\bullet$ Conditional (both pos and neg) & \\
                  & $\bullet$ Default rules         & \\
                  & $\bullet$ A-rewriting (lists)   & \\  
\hline
Clean             & Functional language             & Native code via \\
\cite{PlasmeijerVanEekelen:94}  & $\bullet$ Higher-order  & ABC abstract \\
\cite{SmetsersEtAL:91}  & $\bullet$ Lazy                  & graph rewriting \\
                  & $\bullet$ Strictness annotations & machine \\
                  & $\bullet$ Polymorphic typing     & \\
\hline
Elan              & Rewriting logic language           & C \\
\cite{MoreauKirchner:98}  & $\bullet$ First-order            & \\ 
                  & $\bullet$ Strategy specification   & \\
                  & $\bullet$ AC-rewriting           & \\
\hline
Haskell           & Functional language              & C \\
\cite{PeytonJoneEtAl:93} & $\bullet$ Higher-order           & \\
\cite{PeytonJones:96}    & $\bullet$ Lazy                   & \\
                  & $\bullet$ Strictness annotations & \\
                  & $\bullet$ Polymorphic typing     & \\
\hline
Opal              & Algebraic programming language   & C \\
\cite{DidrichEtAl:94} & $\bullet$ Higher-order       & \\
                  & $\bullet$ Strict                  & \\
\hline
SML               & Functional language              & Native code \\
\cite{Appel:92}   & $\bullet$ Higher-order           & \\
                  & $\bullet$ Strict                  & \\
                  & $\bullet$ Polymorphic typing     & \\
\hline
\end{tabular}
\end{minipage}
\caption{\label{table:LANGUAGES} Languages used in the benchmarking of the \asfsdf\ compiler.}
\end{table}

Table~\ref{table:LANGUAGES} lists some of the semantic features of the
languages used in the benchmarking of the \asfsdf\ compiler.
Modularization aspects are not included.  Although the languages
listed are all based on some form of rewriting, their authors do not
use the same terminology to classify them as can be seen in the second
column.  At least to some extent, this reflects a difference in
orientation and purpose. 

Section~\ref{sec:SmallBenchmarks} gives results of three benchmarks
comparing the compilers for the languages listed in
Table~\ref{table:LANGUAGES}.  Section~\ref{sec:BigBenchmarks} gives
results for two large \asfsdf\ definitions.

\subsection{Three Small Benchmarks} \label{sec:SmallBenchmarks} 

All three benchmarks are based on the normalization of expressions ${2
^ n} \bmod 17$, with $17 \leq n \leq 23$, where the natural numbers
involved are in successor representation (unary representation). They
are synthetic benchmarks yielding rewrite intensive computations.  The
fact that there are much more efficient ways to compute these
expressions is of no concern here, except that this makes it easy to
validate the results.  The sources are available in \cite{Olivier99}.

Note that these benchmarks were primarily designed to evaluate
specific implementation aspects, such as the effect of subterm
sharing, lazy evaluation, and the like. They do not provide an overall
comparison of the various systems.  Also note that some systems
failed to compute results for the full range $17 \leq n
\leq 23$. In those cases, the corresponding graph
ends prematurely.  The possibility to switch subterm sharing off was
added to the \asfsdf\ compiler only for the purpose of
benchmarking. It is not a standard compiler option.  Measurements were
performed on a SUN ULTRA SPARC-5 (270 MHz) with 512 MB of memory.

\subsubsection{The {\tt evalsym} Benchmark}

\begin{figure}
  \caption{\label{evalsym-tlin} Execution times for the {\tt evalsym} benchmark}
\centerline{\psfig{figure=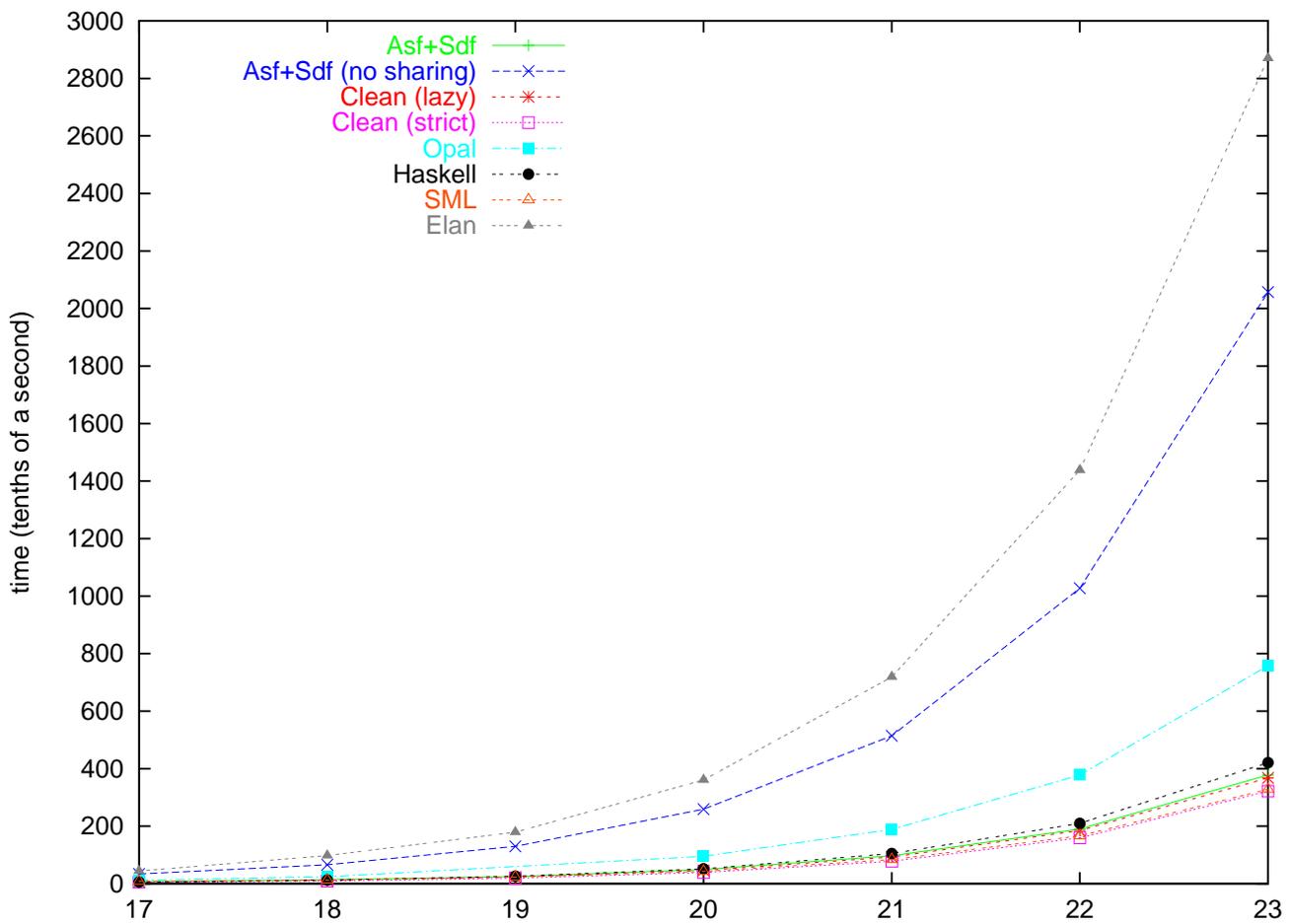,width=12.5cm}}
\end{figure}

The first benchmark is called {\tt evalsym} and uses an algorithm that
is CPU intensive, but does not use a lot of memory.  The results are
shown in Fig.~\ref{evalsym-tlin}.  The differences between
\asfsdf, Clean, Haskell, and SML are small. Even in this case,
maximal subterm sharing is effective in the sense that
\asfsdf\ without sharing performs less well, largely as
a consequence of the less efficient evaluation of {\tt term\_equal}
(Sec.~\ref{sec:TermStorage}), but it does not yield a speed-up with
respect to Clean, Haskell, and SML.  This shows maximal subterm
sharing to be an effective substitute for the sophisticated
optimization techniques used by some of the other compilers. This is
further confirmed by the following two benchmarks.

\begin{figure}
  \caption{\label{evalexp-mlin} Memory usage for the {\tt evalexp} benchmark}
\centerline{\psfig{figure=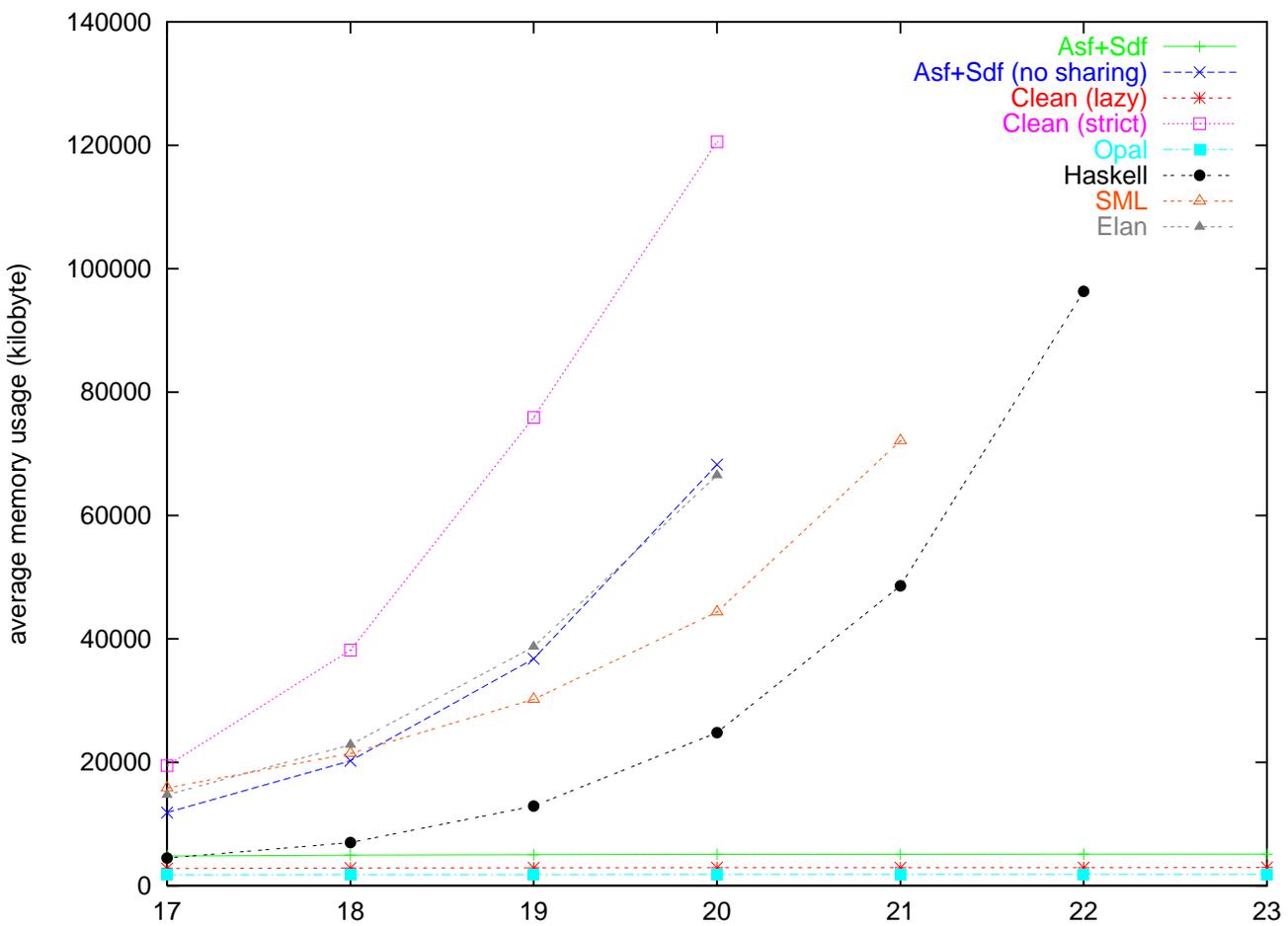,width=12.5cm}}
\end{figure}

\begin{figure}
  \caption{\label{evalexp-tlin} Execution times for the {\tt evalexp} benchmark}
\centerline{\psfig{figure=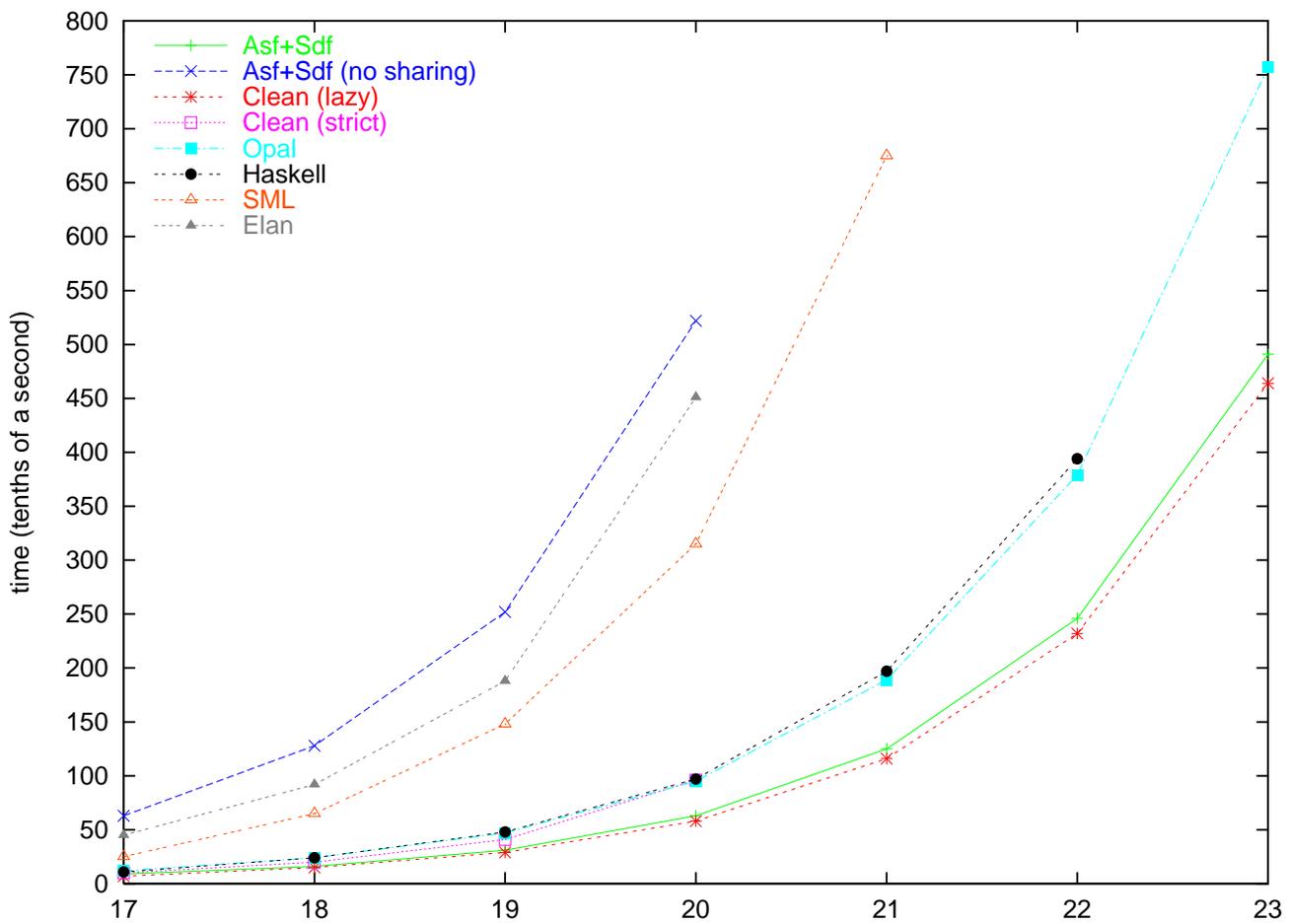,width=12.5cm}}
\end{figure}

\begin{figure}
  \caption{\label{evaltree-mlin} Memory usage for the {\tt evaltree} benchmark}
\centerline{\psfig{figure=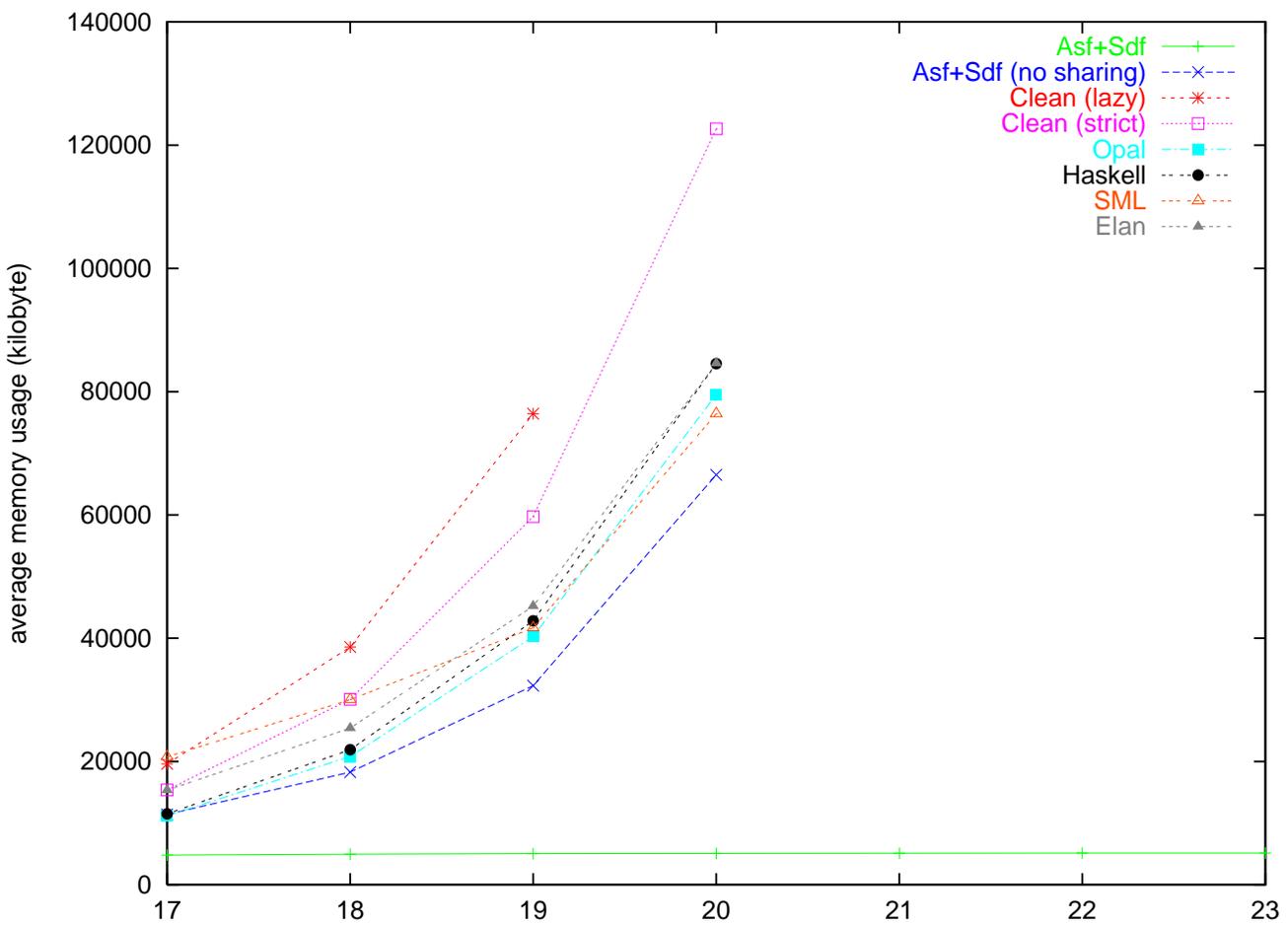,width=12.5cm}}
\end{figure}

\begin{figure}
  \caption{\label{evaltree-tlin} Execution times for the {\tt evaltree} benchmark}
\centerline{\psfig{figure=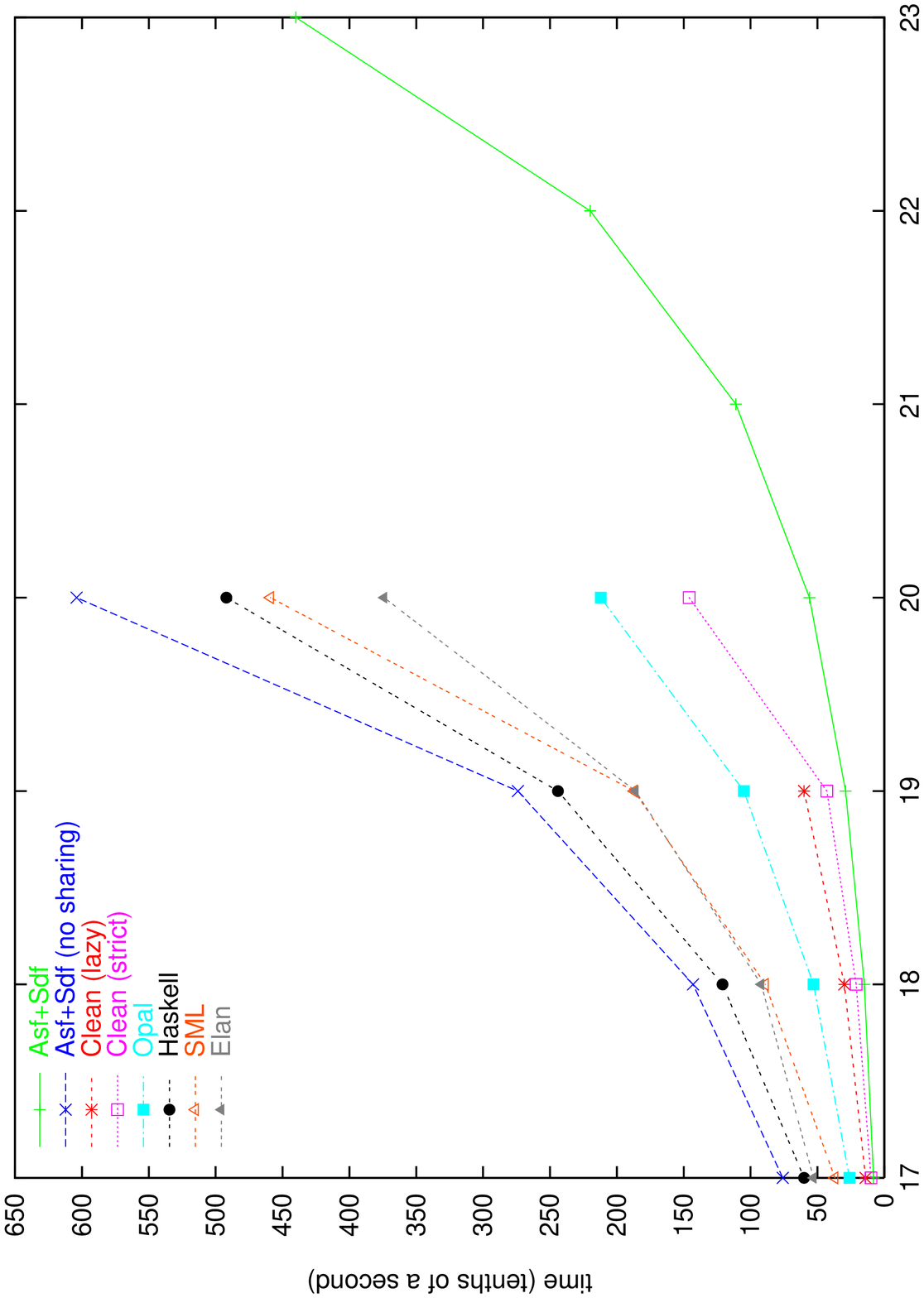,width=12.5cm}}
\end{figure}

\subsubsection{The {\tt evalexp} Benchmark}

The second benchmark is called {\tt evalexp} and is based on an
algorithm that uses a lot of memory when a typical strict
implementation is used. Using a lazy implementation, the amount of
memory needed is relatively small.

Memory usage is shown in Figure~\ref{evalexp-mlin}.  Clearly, strict
implementations that do not use maximal subterm sharing cannot cope
with the excessive memory requirements of this benchmark, but \asfsdf\
and Clean (lazy) have no problems whatsoever.

Execution times are plotted in Figure~\ref{evalexp-tlin}. Only Clean
(lazy) is faster than \asfsdf, but the differences are small.

\subsubsection{The {\tt evaltree} Benchmark}

The third benchmark is called {\tt evaltree} and is based on an
algorithm that uses a lot of memory both with lazy and strict
implementations.  Figure~\ref{evaltree-mlin} shows that neither the
lazy nor the strict implementations can cope with the memory
requirements of this benchmark.  \asfsdf\ is the only one that scales
up for $n > 20$. It can keep memory requirements at an acceptable
level due to its maximal subterm sharing.  The execution times are
shown in Figure~\ref{evaltree-tlin}.

\subsection{Two Large \asfsdf\ Definitions} \label{sec:BigBenchmarks} 

Table~\ref{table:LargeDefs} gives some statistics for two large
\asfsdf\ definitions whose performance is shown in Table~\ref{table:LargeDefsPerf}. 
The \asfsdf\ compiler was written in \asfsdf\
itself, so the top entry in the fourth column of Table~\ref{table:LargeDefs} gives the
self-compilation time.  The language Risla
is a domain-specific language for loans, mortgages,
and other financial products offered by banks \cite{BDKKM96,DK98}. The
expander is the first phase of the Risla implementation. It brings
Risla specifications in normal form by eliminating their modular
structure (if any) \cite{ADR95}.  The C compilation times in the last
column were obtained using SUN's native C compiler with maximal
optimizations.

Table~\ref{table:LargeDefsPerf} gives performance figures for the
compiled versions both with and without maximal subterm sharing of
ATerms (Sec.~\ref{sec:TermStorage}). The time obtained for the
\asfsdf\ compiler with sharing is, of course, again the self-compilation time.

\begin{table}
\begin{center}
\begin{tabular}{|l|c|c|c|c|c|} \hline
Definition & \asfsdf\      & \asfsdf\       & Generated   	& \asfsdf\ to C  & C \\
	      & (rules)       & (lines)     & C code  	        & compilation    & compilation \\
              &               &             & (lines)           & time (s)       & time (s)  \\ \hline \hline
\asfsdf\ compiler          & 1876     & 8699     & 85185    	& 216  		 & 323 \\ \hline
Risla expander             & 1082     & 7169     & 46787    	& 168  		 & 531 \\ \hline
\end{tabular}
\vspace{\baselineskip}
\caption{Size and compilation time for two large \asfsdf\ definitions.} \label{table:LargeDefs} 
\end{center}
\end{table}

\begin{table}
\begin{center}
\begin{tabular}{|l|c|c|} \hline
Application 				& Time (s) 	& Memory (MB) \\ \hline \hline
\asfsdf\ compiler (with sharing)    	& 216		& \ 16	 \\ \hline
\asfsdf\ compiler (without sharing) 	& 661  		& 117 	 \\ \hline
Risla expansion (with sharing)   	& \ \ 9  	& \ \ 8  \\ \hline
Risla expansion (without sharing)  	& \ 18 		& \ 13	 \\ \hline
\end{tabular}
\vspace{\baselineskip}
\caption{Performance of two large \asfsdf\ definitions with and without maximal subterm sharing.} \label{table:LargeDefsPerf}
\end{center}
\end{table}

\section{Conclusions and Further Work} \label{sec:CONC}

The \asfsdf\ compiler generates high quality C code in a relatively
straightforward way. The main factors contributing to its performance
are the decisions to generate C code directly and to use a run-time
term storage scheme based on maximal subterm sharing.  Some
possibilities for further improvement and extension are:
\begin{itemize}

\item Incorporation of additional preprocessing steps such as
      argument reordering during matching, evaluation of sufficiently
      simple conditions during matching in a dataflow fashion, i.e.,
      as soon as the required values become available, and reordering
      of independent conditions.

\item Optimization of repeated applications of a rule like
      rule~{\tt [s-1]} in Sec.~\ref{sec:listmatching}, or of
      successive applications of different rules by analyzing their
      left- and right-hand sides. Similarly, elimination of the redex
      search phase in some cases (``matchless rewriting'').

\item Incorporation of other rewrite strategy options besides
      default rules and the {\tt delay} attribute that are currently
      supported.

\end{itemize}

\noindent \begin{acks}
We would like to thank Hayco de Jong for his contribution to the
implementation of the ATerm library, Jurgen Vinju for looking into the
efficiency of list matching, Wan Fokkink for his useful remarks, and
Pierre-Etienne Moreau for discussions on the compilation of term
rewriting systems in general.  The idea for the benchmark programs in
Section~\ref{sec:SmallBenchmarks} is due to Jan Bergstra.
\end{acks}

% The following two commands are all you need in your .tex file to
% produce the bibliography for the citations herein.  
% You must have a proper .bib file, and remember to run:
% latex bibtex latex latex
% to resolve all references
%
\bibliographystyle{acmtrans}

% \bibliography{/ufs/jan/ASF+SDF/compiler/compiler,/ufs/jan/bibfiles/LanguageProto,/ufs/jan/bibfiles/metabib}

% HOWEVER, having run 'latex bibtex latex latex' and thus, having
% obtained your .bbl file, you should read it into your main
% file (at the end) just like this:
%

\end{thebibliography}

\end{document}